\documentclass[%
twocolumn,
 amsmath,amssymb,
prl
floatfix,
]{revtex4-2}

\usepackage{graphicx}
\usepackage{dcolumn}
\usepackage{bm}
\usepackage{amsmath}
\usepackage{comment}

\usepackage{tabularx}
\usepackage{bm}
\usepackage{euscript}
\usepackage{epsfig,psfrag,subfigure}
\usepackage{graphicx}
\usepackage{color}
\usepackage{amsfonts}
\usepackage{exscale}
\usepackage{amsbsy}
\usepackage{multirow}
\usepackage[normalem]{ulem}
\usepackage{stmaryrd}
\usepackage{wasysym}
\usepackage{comment}

\def\abs#1{\left|#1\right|}

\def\be{\begin{equation}}       \def\ee{\end{equation}}
\def\bea{\begin{eqnarray}}      \def\eea{\end{eqnarray}}
\def\ba{\begin{array} }
\def\ea{\end{array} }
\def\bnum{\begin{enumerate} }
\def\enum{\end{enumerate}}

\def\=>{\Rightarrow}
\def\>{\rightarrow}

\def\eye2{Fathbb{I}}

\usepackage[colorlinks,linkcolor=blue,citecolor=blue,anchorcolor=blue]{hyperref}

\begin{document}

\title{Anomalous superfluid density in a disordered charge density wave material: Pd-intercalated ErTe$_3$}

\author{Yusuke Iguchi$^{1,2,3}$, Joshua A. Straquadine$^{2,3}$, Chaitanya Murthy$^{1,4}$, Steven A. Kivelson$^{1,2,4}$, Anisha G. Singh$^{2,3}$, Ian R. Fisher$^{1,2,3}$, and Kathryn A. Moler$^{1,2,3,4}$}
\affiliation{
$^1$Stanford Institute for Materials and Energy Sciences, SLAC National Accelerator Laboratory, 2575 Sand Hill Road, Menlo Park, California 94025, USA\\
$^2$Geballe Laboratory for Advanced Materials, Stanford University, Stanford, California 94305, USA\\
$^3$Department of Applied Physics, Stanford University, Stanford, California 94305, USA\\
$^4$Department of Physics, Stanford University, Stanford, CA 94305, USA
}%

\date{\today}

\begin{abstract}
We image local superfluid density 
in single crystals of Pd-intercalated ErTe$_3$
below  the superconducting critical temperature, $T_c$,
well below the onset temperature, $T_{CDW}$, of (disordered) charge-density-wave order.
We find no detectable inhomogeneities.
We observe a rapid increase of the superfluid density below $T_c$, deviating from the behavior expected in 
conventional Bardeen-Cooper-Schrieffer, and show that the temperature dependence is qualitatively consistent with a combination of quantum and thermal phase fluctuations.
\end{abstract}

\maketitle


Pd$_x$ErTe$_3$ is a model system for quasi-two-dimensional (2D) superconductivity and for the competition between charge-density-wave (CDW) and superconducting (SC) states. The superfluid density characterizes the phase stiffness of the superconducting order parameter and determines the London penetration depth $\lambda(T)$. In a conventional 3D Bardeen-Cooper-Schrieffer (BCS)  superconductor, the temperature dependence of the normalized superfluid density, $n_s(T)=\lambda^2(0)/\lambda^2(T)$, is controlled by the population of thermally excited Bogoliubov quasiparticles, and can be calculated using the Bogoliubov–de Gennes equations~\cite{Kita2015} or the semi-classical model~\cite{Chandrasekhar1993}. At low temperatures, measurements of $n_s(T)$ provide information about the superconducting gap structure $\Delta(T,\mathbf{k})$. 
At temperatures close to $T_c$, however, the same theoretical considerations imply that $dn_s(T)/dT|_{T\rightarrow T_c}$ is not very sensitive to the gap structure, and changes somewhat but not dramatically in the strong-coupling and/or dirty limits~\cite{Xu1995,Maisuradze2012}. 

$n_s(T)$ may have distinct features in quasi-2D conventional BCS superconductors. When the superconducting coherence length $\xi$ is larger than the film thickness, the Berezinskii-Kosterlitz-Thouless (BKT) theory predicts an anomaly in the superfluid density at the BKT transition temperature~\cite{Benfatto2008,BKT1, BKT2}. More generally, strong phase fluctuations may suppress $T_c$ and increase $dn_s(T)/dT|_{T\rightarrow T_c}$~\cite{Emery1995}. Such anomalies have been observed in various ultra-thin film superconductors, including Y$_{1-x}$Ca$_x$Ba$_2$Cu$_3$O$_{7-\delta}$~\cite{Hetel2007}, NbN~\cite{Kamlapure2010}, Pb~\cite{Nam2016}, and $a$-MoGe~\cite{Mandal2020}.

We conducted measurements of the local diamagnetic susceptibility in Pd$_x$ErTe$_3$ ($0<x<0.06$), a quasi-2D layered bulk superconductor, using scanning superconducting quantum interference device (SQUID) microscopy (SSM) with micron-scale spatial resolution. Our results show that the superfluid density is homogeneous, with no detectable heterogeneity on micron scales. Additionally, we find non-BCS-like temperature dependence of the superfluid density with a steep slope $dn_s(T)/dT$ near $T_c$. 

\begin{figure}[b]
\includegraphics*[width=8.5cm]{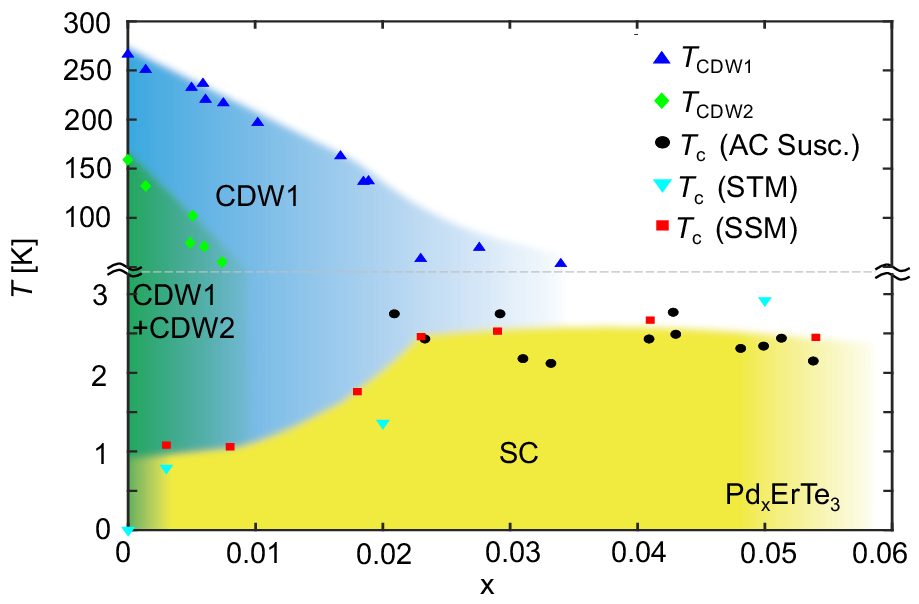}
\caption{\label{fig:PD} Phase diagram of Pd-intercalated ErTe$_3$. $T_{\mbox{\scriptsize CDW1,2}}$ from Ref.~\cite{Josh2019}. $T_c$ determined by bulk ac susceptibility~\cite{Josh2019} and STM~\cite{Alan2020}. $T_c$ obtained in this work (SSM) are plotted as red squares.}
\end{figure}


Recently, intertwined SC and CDW order has been observed in Pd-intercalated ErTe$_3$~\cite{He2016,Josh2019,Alan2020}. The pristine `parent' compound ErTe$_3$ shows two, mutually transverse, in-plane, unidirectional, incommensurate CDW states~\cite{Ru2008}, with no SC down to the measured lowest temperature, 100~mK ~\cite{Alan2020}. Pd-intercalation induces disorder in the crystal lattice, suppressing CDW formation and leading to a SC ground state [Fig.~1]~\cite{Josh2019,Alan2020}.   In crystals with a Pd concentration near $x = 0.05$, long-range CDW is not observed~\cite{Alan2019}. Scanning tunneling microscopy (STM) measurements of the tunneling conductance revealed a homogeneous SC gap at length scales exceeding the SC coherence length, and showed no direct correlation between the CDW and SC orders~\cite{Alan2020}. 
The anisotropic in-plane coherence lengths were estimated as $\xi_a \sim 1500$~\AA~and $\xi_c\sim1000$~\AA ~\cite{Alan2020}.

For this work, bulk single crystals of Pd-intercalated ErTe$_3$ were grown using the flux method~\cite{Josh2019}. We made images with a scanning SQUID susceptometer on cleaved {\it b}-planes of Pd$_x$ErTe$_3$ at temperatures varying from 0.3 K to 3 K in a Bluefors LD dilution refrigerator for samples with $x=.003, .008, .018, .023, .029, .041, .054$. Our scanning SQUID susceptometer has a pickup loop that measures the local magnetic flux $\Phi$ in units of the flux quantum $\Phi_0$~\cite{Kirtleyrsi2016} while scanning with a pickup loop-sample separation $z$, which we call the height. The minimum $z$ can vary slightly between cooldowns and is ~800 nm in these measurements (supplemental material ~\cite{supple}). The pickup loop is paired with a concentric field coil through which we apply an ac current of $|I^{\text{ac}}| =$ 1~mA at a frequency of 1~kHz using an SR830 Lock-in-Amplifier to produce a spatially varying localized ac magnetic field~\cite{Kirtleyrsi2016}. We measure both quasi-static flux and the ac magnetic flux $\Phi^{\text{ac}}$, and report the local ac susceptibility as $\chi=\Phi^{\text{ac}}/|I^{\text{ac}}|$ in units of $\Phi_0$/A. SSM has been employed to image inhomogeneous superfluid responses in unconventional superconductors by detecting the local ac magnetic susceptibility~\cite{Cliff2009,Beena2010,Chris2018,Irene2019,Logan2019,Iguchi2022}. By measuring the dependence of the local susceptibility on the scanning SQUID height, SSM enables estimation of the local London penetration depth $\lambda$~\cite{Cliff2009,Luan2011, Kirtley2012, Lippman2012, Julie2012, Iguchi2021,Iguchi2022}.

\begin{figure}[tb]
\includegraphics*[width=8.5cm]{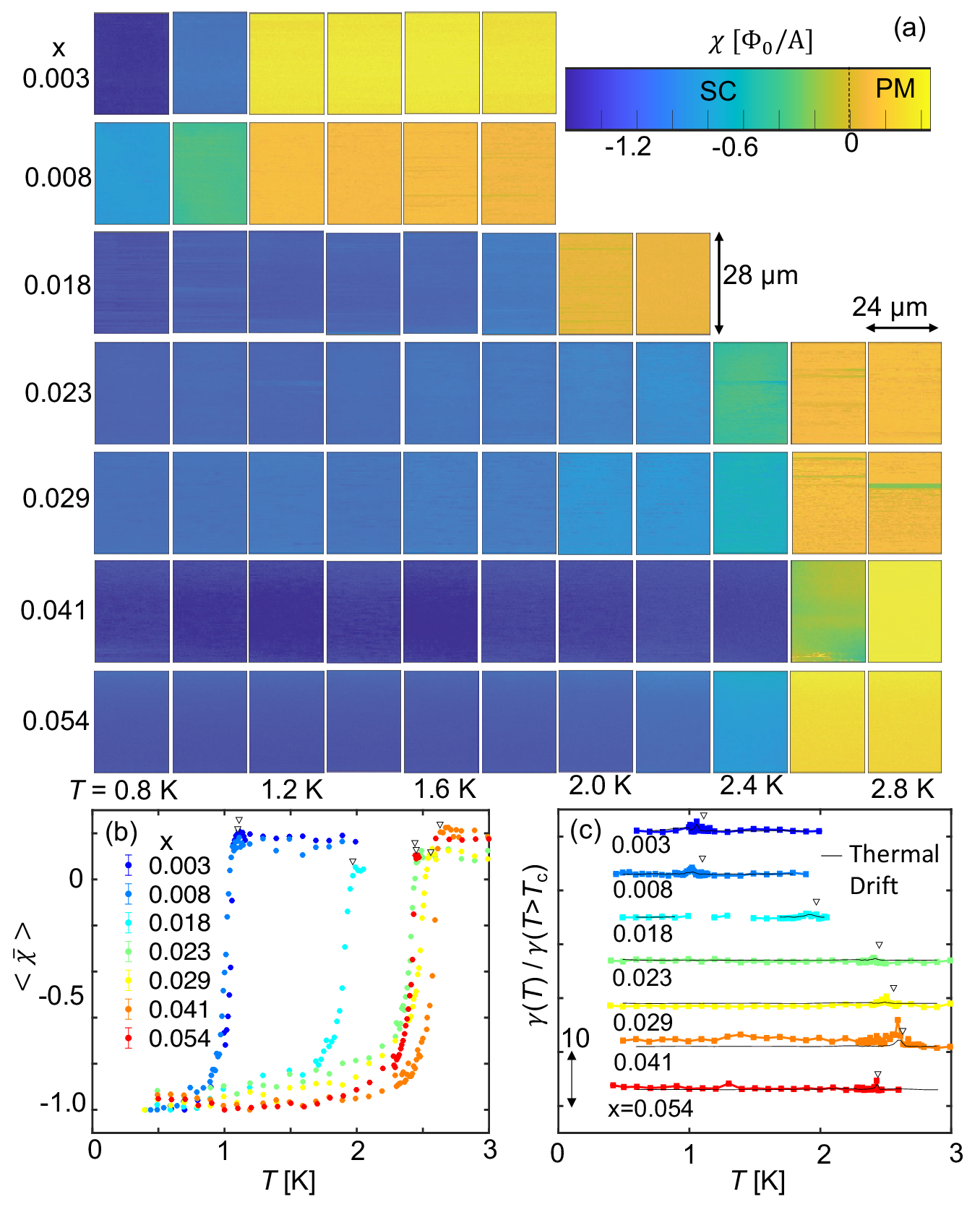}
\caption{\label{fig:Homo} Homogeneous superfluid density on micron scales in Pd$_x$ErTe$_3$. (a) Temperature dependence of local susceptibility images. (b) Normalized average susceptibilities show sharp drops just below $T_c$. (c) The standard deviation of the susceptibility shows only small peaks near $T_c$, consistent with thermal drift. Inverted triangles indicate $T_c$ and solid lines are numerical calculations, including thermal drifting of $\pm$5~mK ~\cite{supple}.}
\end{figure}

To investigate the inhomogeneity of superfluid response, we imaged the local susceptibility at several temperatures. In all samples over the entire range of Pd concentrations explored, we observed sharp and apparently homogeneous transitions from the paramagnetic phase to the SC diamagnetic phase with $T_c$'s in the
range $T_c=0.8-2.8$~K [Fig.~2(a)]. The slight variation in the observed paramagnetic (PM) susceptibility above $T_c$ among different Pd concentrations could represent variation as a function of doping but could also be due to differences in scan heights. 

We analyze the susceptibility images by constructing a histogram of the number of pixels with a given amplitude of $\chi$. The histograms show sharp peaks, indicating a relatively homogeneous sample. The spacing between pixels is 300~nm, and each pixel samples a micron-scale area determined by the geometry of the pickup loop and field coil. We choose a Gaussian function of the form $\mathcal{N} \exp{(-(\chi-\beta)^2/2\gamma^2)}$ to fit the peaks in the histogram~[Supplemental Fig.~S1]~\cite{supple}. The normalized susceptibility averaged over the image is $<\bar{\chi}> \equiv \beta(T)/\beta$(0.5~K), and the upper limit on the inhomogeneity of the superfluid response on micron scales is characterized by the normalized standard deviation, $\gamma(T)/\gamma(T>T_c)$. Plotting $<\bar{\chi}>$ vs. $T$, we see that $T_c$ as a function of Pd concentration [Fig.~2(b)] is consistent with previous measurements based on bulk susceptibility and STM measurements~\cite{Josh2019,Alan2020}. The upper limits on the inhomogeneity exhibit small peaks just below $T_c$ [Fig.~2(c)] and are consistent with a slight thermal drift during the scan [Supplemental Fig.~S2]~\cite{supple}. Thus, the superfluid response in Pd$_x$ErTe$_3$ ($x$=0.003-0.054) is consistent with homogeneity on a micron scale.

\begin{figure}[hb]
\includegraphics*[width=8cm]{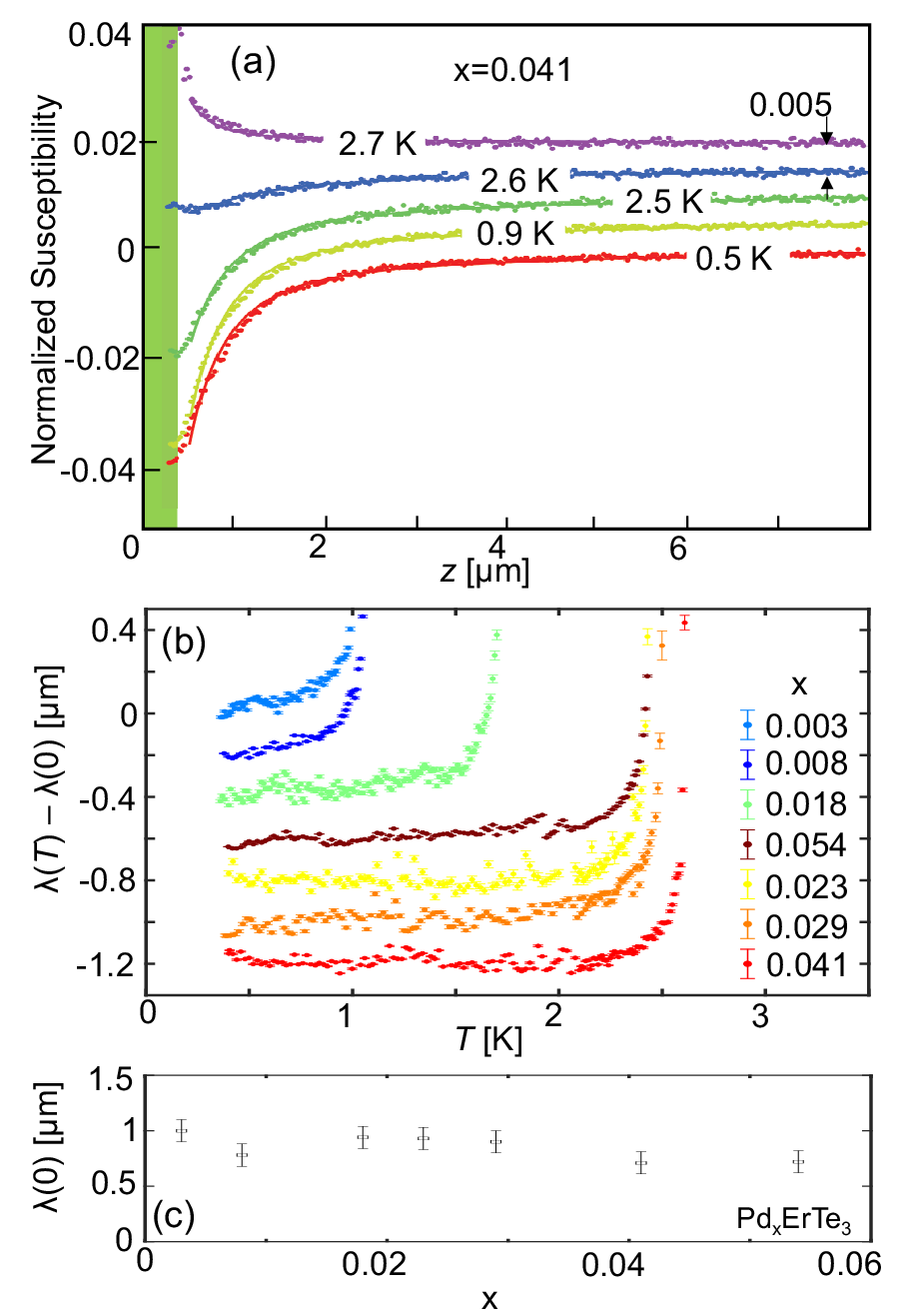}
\caption{\label{fig:Td} Local susceptibility vs. height provides $\lambda(T)$. (a) Height dependence of local normalized susceptibility $\chi(z)/\phi_s$ in $x = 0.041$ sample is well fitted by numerically calculated curves (solid lines) using Supplemental Eq.~(S4) with $\lambda(T)$ as a fitting parameter. The green-color-filled area indicates the distance between the pickup loop's center and the sample surface when the SQUID tip touches the surface~\cite{supple}. (b) Temperature dependence of the penetration depth obtained from the fitting results of Fig.~2(a) are plotted with an offset of 200~nm. (c) Estimated penetration depth at $T = 0$.}
\end{figure}

To determine the penetration depth, we measured susceptibility vs. height [Fig.~3(a)]. The susceptibility is paramagnetic above $T_c$ and diamagnetic below $T_c$. We fit the height dependence of the susceptibility~\cite{supple} to a model that assumes a circular pickup loop of radius $r'$ and field coil of radius $r$ at a height $z$ above the top of a thin film of thickness $t$ on a substrate. The thin film is characterized by a London penetration depth $\lambda$ and a paramagnetic permeability $\mu_2$. We estimate the permeability $\mu_2=1.03\mu_0$, where $\mu_0$ is the permeability of vacuum, by fitting the height dependence of the paramagnetic susceptibility above $T>T_c$ to Supplemental Eq.~(S4) with fixed parameters $t$,$r'$,$r$, and free parameter $\mu_2$. We then estimate $\lambda(T)$ by fitting susceptibility vs. $z$ for each value of $T<T_c$ to Supplemental Eq.~(S4) with fixed parameters $t$,$r'$,$r$, $\mu_2$, a copper substrate permeability $\mu_3=\mu_0$, and free parameter $\lambda(T)$.

The penetration depth does not depend strongly on temperature at low temperatures [Fig.~3(b)]. We estimate $\lambda(T=0)$ across the doping series to be in the range of 700-1000~nm, consistent with measurements of an isolated vortex field [Supplemental Fig.~S3]~\cite{supple}. 
This penetration depth is a factor of $3.5-5$ larger than the only other estimate of $\lambda$ in this material of which we are aware, which was an indirect estimate from the lower critical magnetic field at $T/T_c \sim 0.7$ for an $x=0.043$ sample~\cite{Alan2020}. The error bars shown in the figure include all sources of error of which we are aware (supplemental materials~\cite{supple}). 
Interestingly, we did not observe a significant dependence of $\lambda(T=0)$ on the Pd-intercalation concentration [Fig.~3(c)]. In BCS theory, $\lambda^2$ would be expected to decrease in proportion to the mean free path ~\cite{Tinkham}, so the flat dependence of $\lambda(0)$ on $x$ suggests either that BCS theory does not apply or that $x$ is not the main determining factor for the mean free path.

Using the obtained values of $\lambda$, we calculate the normalized superfluid density $n_s(T) = \lambda^2(0)/\lambda^2(T)$. Our results reveal a rapid increase of $n_s$ with decreasing temperature just below $T_c$ and a slower increase at lower temperatures [Fig.~4(a)]. This temperature dependence clearly deviates from the expectations of the conventional weak coupling {\it s}-wave model (BCS model).


To investigate whether the anomalous temperature dependence of $n_s$ can be simply attributed to details of the gap structure or strong coupling effects, we consider an anisotropic {\it s}-wave model. In this model, the superconducting gap is described as $\Delta(T,{\mathbf{k}}) = \Delta_0(T)\times g({\mathbf{k}})$, where $\Delta_0(T)$ represents the temperature dependence of the gap, and $g({\mathbf{k}})$ its angular variation on the Fermi surface~\cite{Prozorov2006}. 
The temperature dependence is approximated by the typical mean-field form $\Delta_0(T) = \Delta_0(0)\tanh(\pi T_c\sqrt{\alpha (T_c/T-1)}/\Delta_0(0))$, where $\Delta_0(0)$ is the gap magnitude at $T=0$ and $\alpha$ is a parameter. For a gap with anisotropic {\it s}-wave symmetry on a 2D cylindrical Fermi surface, $g(\phi) = \sqrt{1-\varepsilon\sin^2\phi}$, where $\phi = 0$ and $\pi/2$ correspond to the {\it a} and {\it c} axes, respectively, and $\varepsilon = 1-\left[\Delta_c(0)/\Delta_a(0)\right]^2$ (assuming that $0<\Delta_c \leq \Delta_a$). 
We note that our model does not determine which axis has a larger gap amplitude, as we take an angular average for the normalized superfluid density. The fitting parameters in this model are $\Delta_0(0)$, $\varepsilon$ and $\alpha$~\cite{supple}, and the normalized superfluid density is:
\begin{multline}
    n_i(T) = 1 - \frac{1}{2\pi T} \int_0^{2\pi} d\phi \, P_i(\phi) \\
    \int_0^\infty d\epsilon \, \cosh^{-2}\!\left(\frac{\sqrt{\epsilon^2+\Delta^2(T,\phi)}}{2T}\right),
\end{multline}
where $i=a,c$, and $P_a = \cos^2\phi$, $P_c=\sin^2\phi$. 
We find that our measured normalized superfluid density $n_s\simeq(n_a+n_c)/2$ can indeed be well fitted using Eq.~(1) (for details of the fits, see the supplemental material~\cite{supple}) [Fig.~4]. However, the fitted parameter $\alpha\sim 10$ is much larger than known models, such as $\alpha=1$ (isotropic {\it s}-wave) and $\alpha=2$ ({\it s}+{\it g}-wave)~\cite{Prozorov2006}. Moreover, the quality of the fit strongly depends on the value of $\alpha$ rather than the anisotropy $\varepsilon$ or the coupling constant $\Delta_0(0)/k_B T_c$. Thus our fitting results suggest that the temperature-dependent superfluid density {\it cannot} fit the BCS model.

\begin{figure}[tb]
\includegraphics[width=8.5cm]{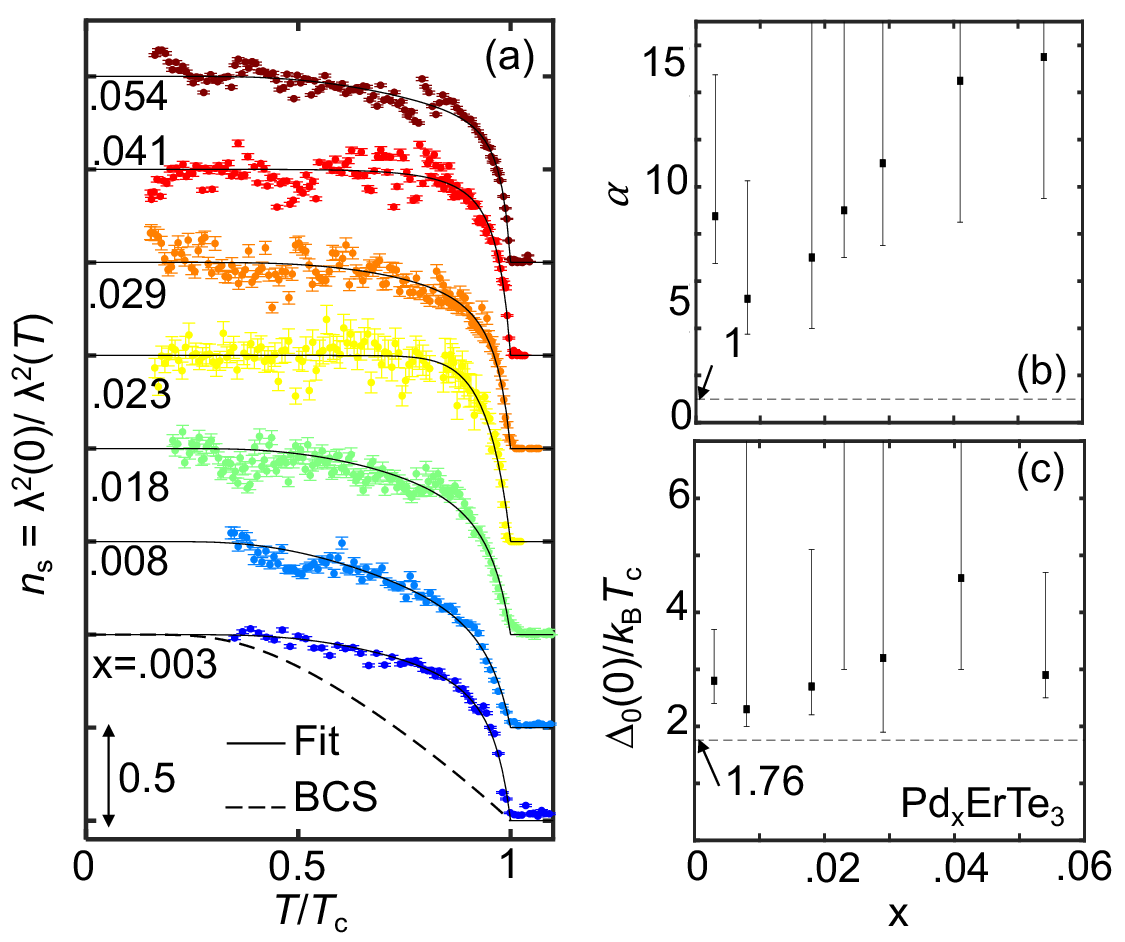}
\caption{\label{fig:AnisoS} Comparison of the estimated normalized superfluid density from Fig. 3(b) to an anisotropic {\it s}-wave BCS model. (a) Superfluid density (dots) and fits (solid lines), offset by 0.5. (b) Fitted values of $\alpha$ {\it vs} $x$. Values $\alpha>>1$ are physically unrealistic for known BCS models. (c) Fitted coupling constant $\Delta_0(0)/k_BT_c$ {\it vs} $x$.}
\end{figure}

\begin{figure}[htb]
\includegraphics[width=8.5cm]{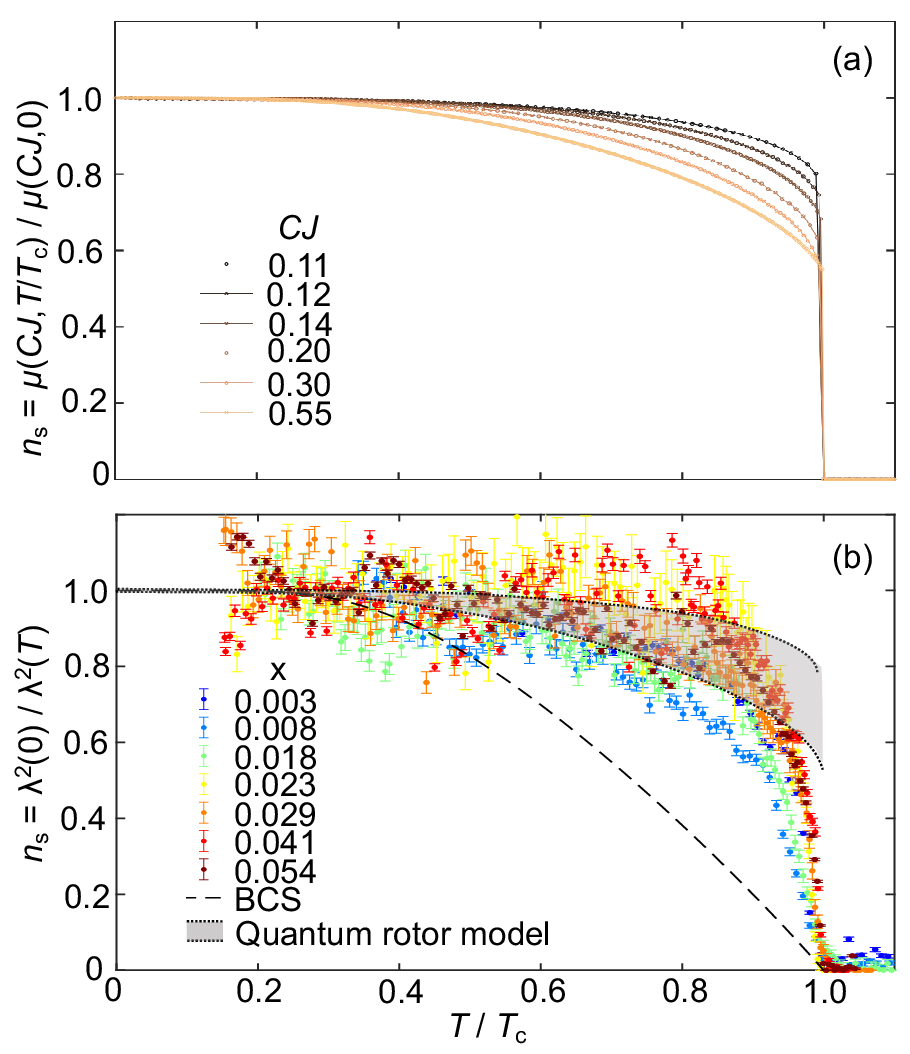}
\caption{\label{fig:QPF} Normalized superfluid density in the quantum rotor model compared to experimental results in Pd$_x$ErTe$_3$. (a) Variational solution of the quantum rotor model for several values of the coupling constant $CJ$. (b) The experimental results from Fig.~4(a) compared with the results of the quantum rotor model from Fig.~5(a) and Supplemental Fig.~S5.}
\end{figure}

We next consider fluctuations, which can suppress $T_c$ and modify the temperature dependence of the superfluid density. Quasi-2D electronic structures can enhance fluctuations ~\cite{Alan2020,Ru2008}. A pure BKT scenario cannot be applied here, as the sample thicknesses exceed the coherence length. Classical phase fluctuations alone would destroy the SC order above $T_\theta =$ 7-14 K, estimated from formulas in Ref. \cite{Emery1995} using $\xi$ = 100-150 nm and $\lambda$ = 700-1000 nm. 
Notably, this estimated $T_\theta$ is close to $T_c$, suggesting that such phase fluctuations might significantly contribute to the determination of $T_c$.
(Note that Fang {\it et al.} estimated $T_\theta$ as 170 K, much larger than $T_c$, from $\lambda$ = 200 nm~\cite{Alan2020}.) However, superfluid density that is dominated by classical phase fluctuations 
would exhibit a linear-$T$ dependence 
well below $T_c$~\cite{Carlson1999}, not flattening until quantum effects become important. 
Therefore, classical phase fluctuations alone cannot explain our results. 

Quantum phase fluctuations may modify this scenario.  The small value of $T_\theta$ and the quasi-2D character of the electronic structure likely enhance the effectiveness of these fluctuations, which may be further enhanced~\cite{Nakhmedov2012} by a degree of randomness 
of the inter-layer Josephson coupling
produced by the Pd intercalation. 
To determine whether a combination of quantum and classical phase fluctuations might account for the observed anomalous $T$-dependence of the superfluid density, 
we have studied a caricature of the problem in terms of the quantum rotor model, governed by the Hamiltonian
\begin{equation}
    H = \sum_{j}{\frac{n_j^2}{2C}} - J\sum_{<i,j>}{\cos{(\theta_i-\theta_j)}},
\end{equation}
where $n_j$ is the number of Cooper pairs on site $j$ and satisfies the commutation relations $\left[ n_i,n_j\right] =\left [e^{i\theta_i},e^{i\theta_j}\right]=0$ and $\left[n_i,e^{i\theta_j}\right] = \delta_{ij} \, e^{i\theta_j}$, $C$ is 
a local capacitance which plays the role of an effective mass, and $J$ is a 
measure of the phase stiffness within each plane. 
(This model omits many possibly significant effects, including long-range Coulomb interactions and dissipation stemming from the existence of quasiparticle excitations.)
For this model, we estimate the $T$-dependent superfluid density 
using the variational method 
used in~\cite{Chakravarty1986} (for details, see the supplemental material~\cite{supple}). 
The 
results for 
a range of coupling constants $C$ and $J$ 
capture some of the salient features of our experimental findings, as shown in Fig.~5(b), suggesting that strong quantum phase fluctuation 
are probably significant.

Finally, it is worth noting that $T_c$ displays a complex variation with $x$ as shown in Fig.~1, where $T_c$ initially rises rapidly with $x$ before approximately ‘saturating’. The fact that $T_c$ does not decrease with the disorder at $x>0.02$ might be attributed to Anderson's theorem, but this theorem does not explain the initial rise relative to zero Pd concentration~\cite{Anderson1959}. The $x$-dependence of $T_c$ likely reflects the complex interplay of a variety of factors, including the competition between CDW formation and superconductivity, the effects of disorder on the CDW state, and also the influence of quantum phase fluctuations on the superconducting state.

In summary, we have used scanning SQUID susceptometry to examine, at the microscopic level, the superfluid response 
on cleaved surfaces of Pd-intercalated ErTe$_3$. 
Our findings reveal that the superfluid response is uniformly  
on a micron scale within the Pd-intercalation-induced superconducting state, consistent with previous STM measurements. We also observe an unexpectedly strong (relative to BCS) temperature dependence of the superfluid density 
near $T_c$ for all Pd concentrations. 
To explain this non-BCS-like temperature-dependent superfluid density in Pd$_x$ErTe$_3$, we employ the quantum rotor model. Our results suggest that quantum phase fluctuations suppress $T_c$ and determine the functional form of $\lambda(T)$ in Pd$_x$ErTe$_3$. Moreover, our study highlights the potential of temperature-dependent superfluid density as a valuable tool for investigating 
quantum phase fluctuations in quasi-2D superconductors.

\begin{acknowledgments}
The authors thank A. Kapitulnik, A. Fang, and J. R. Kirtley for fruitful discussion. This work was primarily supported by the Department of Energy, Office of Science, Basic Energy Sciences, Materials Sciences and Engineering Division, under Contract No. DE- AC02-76SF00515. Y.I. was partially supported by a JSPS Overseas Research Fellowship.
C.M. was supported in part by the Gordon and Betty Moore Foundation's EPiQS Initiative through GBMF8686.
\end{acknowledgments}


\newpage
\clearpage

\setcounter{figure}{0}
\setcounter{equation}{0}
\renewcommand{\thefigure}{S\arabic{figure}}
\renewcommand{\theequation}{S\arabic{equation}}

\onecolumngrid
	
\begin{center}
	\Large
	{Supplemental Material for \\\lq\lq Anomalous superfluid density in a disordered charge density wave material: Pd-intercalated ErTe$_3$ \rq\rq} \\by Iguchi $et$ $al.$
\end{center}

\section{Inhomogeneity of superfluid response}

\begin{figure}[htb]
\includegraphics*[width=8cm]{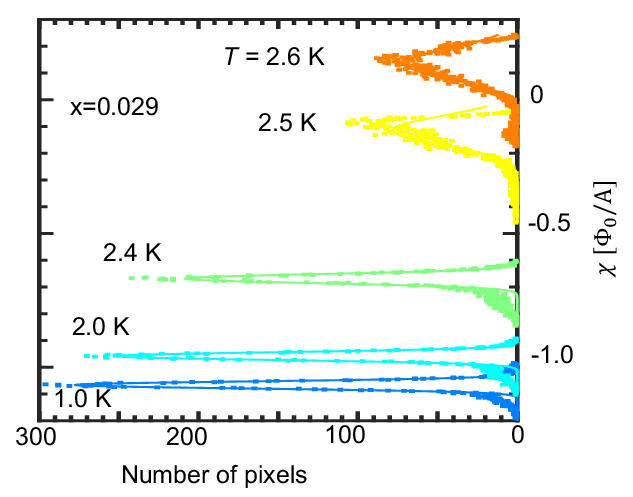}
\caption{\label{fig:inhomo} Number of pixels showing the same susceptibility at $T$ = 1.0-2.6 K in x = 0.029. Solid lines are fitting curves of $\mathcal{N} \exp(- (x-\beta)^2/2\gamma^2)$.}
\end{figure}

We estimated the noise in the inhomogeneity $\gamma(T)/\gamma(>T_{\mbox{\scriptsize c}})$ due to the experimental noise of susceptibility measurement and the thermal drift effect. The experimental noise of susceptibility, $\chi_{N} = \pm$ 0.025~$\Phi_0$/A, does not depend on the temperature, thus this noise is considered as $C_{\mbox{\scriptsize ex}} = 1$. If we assumed the drifting temperature range was $\Delta T_{N} = \pm$5~mK, the noise of the thermal drift effect was estimated as,
\begin{equation}
    C_{\mbox{\scriptsize th}} = (\Delta\chi(T_{m+1})+\Delta\chi(T_m))\times\Delta T_{N},
\end{equation}
where 
\begin{equation}
    \Delta\chi(T_m)= \frac{\chi(T_{m+1})-\chi(T_m)}{T_{m+1}-T_m}\frac{1}{\bar{\chi}_N},
\end{equation}
\begin{equation}
    \bar{\chi}_N = \frac{\chi_N}{\beta(0.5 \mathrm{K})}.
\end{equation}
The estimated total noises of $C_{\mbox{\scriptsize tot}} = C_{\mbox{\scriptsize ex}} + C_{\mbox{\scriptsize th}}$ were quantitatively consistent with the observed data in Fig. 2(c).

\begin{figure*}[htb]
\includegraphics*[width=14cm]{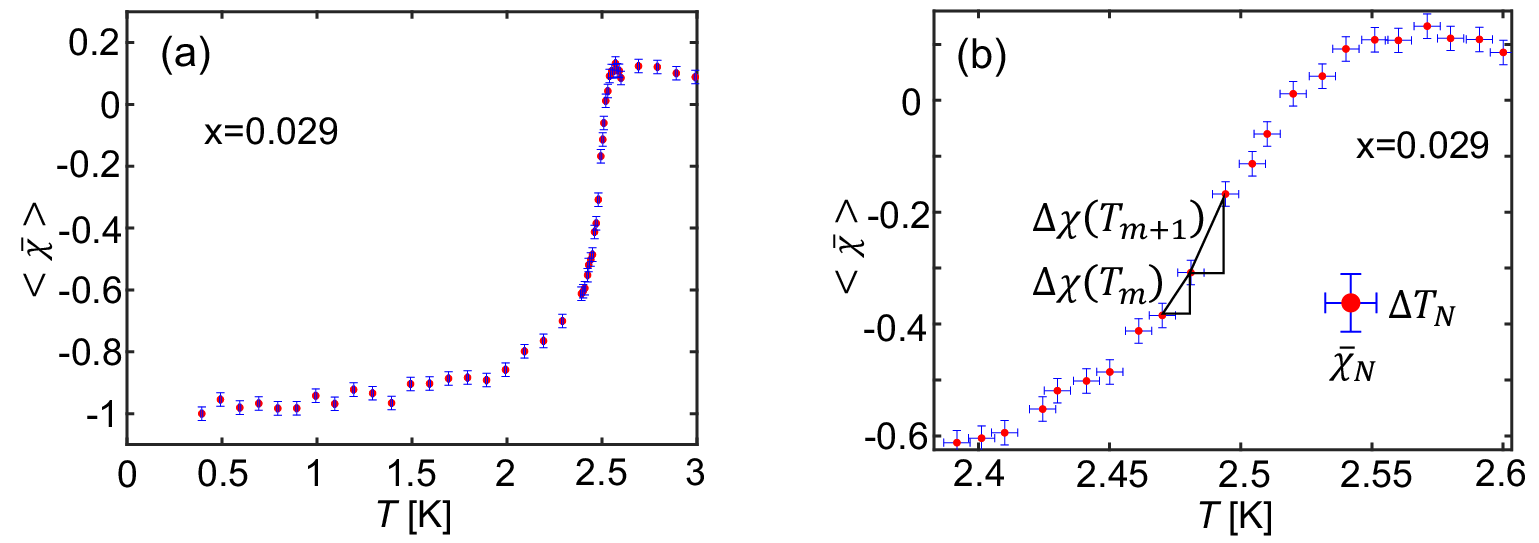}
\caption{\label{fig:noise} (a) Normalized averaged susceptibility with experimental noise of $\pm0.025~\Phi_0$/A. (b) Normalized averaged susceptibility with experimental noise of $\pm0.025~\Phi_0$/A and thermal drift of $\pm5$~mK.}
\end{figure*}

\clearpage
\section{Isolated vortex field}
We observed an isolated vortex at $x = 0.29$ at 0.5~K, which was shown in Fig. A3. The cross section of the magnetic flux was consistent with the numerical simulation of a point source magnetic monopole field with total flux $\Phi=\Phi_0$, which includes the SQUID structure\cite{Kirtleysst2016}. In this model, the magnetic monopole is set at $z = 0$, where the sample surface is at $z = \lambda$, and and the center of pickup loop is at $z=z_0+\lambda$, where $z_0 = 800$ nm and the magnetic monopole field is defined as $H(\vec r) = \Phi_0z/\mu_0r^3$. The best fit curve used $\lambda$ = 1.0~$\mu$m ($\pm$0.2~$\mu$m). This is consistent with the result from the susceptibility height dependence(Fig. 3(c)).

\begin{figure}[htb]
\includegraphics*[width=8cm]{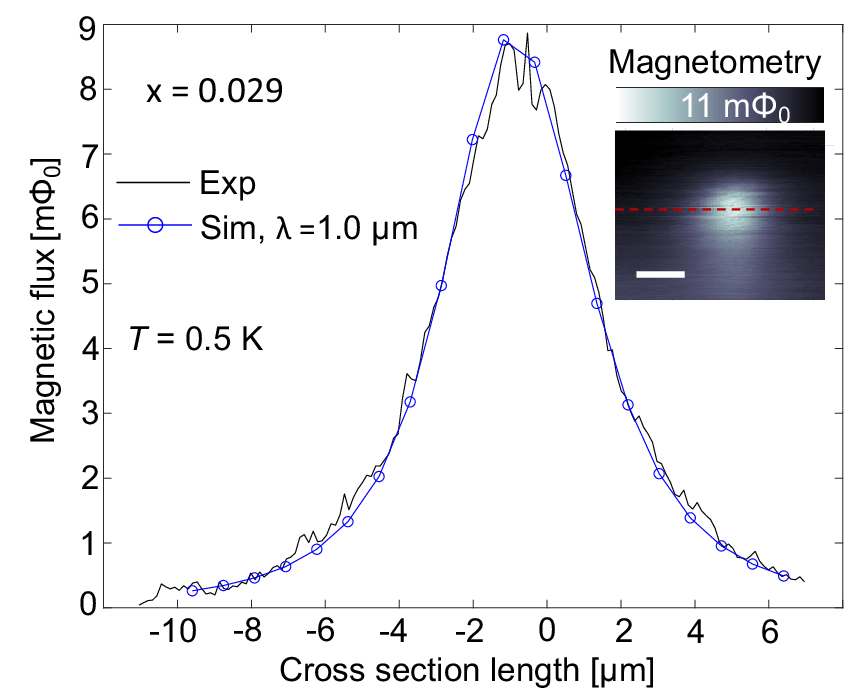}
\caption{\label{fig:vortex} Cross section of an isolated vortex field observed was fitted by the point source model. In the inset, the dashed line shows the cross section and scale bar is 5 $\mu$m.}
\end{figure}

\clearpage

\section{Estimate the London penetration from Susceptibility measurements}

To estimate the local $\lambda$, we fit the observed height dependence of $\chi$ to an expression~[Eq.~(S4)] developed by Kirtley {\it et al.} for the case of homogeneous and isotropic $\lambda$~\cite{Kirtley2012}. This approach is valid if $\lambda$ varies slowly on the relevant length scales and is approximately isotropic in-plane (the out-of-plane $\lambda$ does not appear in the result for $\chi$ in this case~\cite{Kogan2003}).
In the model, the sample surface is the $z=0$ plane. The pickup loop and field coil are at $z>0$ in vacuum, where the permeability is $\mu_0$. In the sample ($0 \geq z \geq -t$, where $t$ is the sample thickness), the London penetration depth is $\lambda$ and the permeability is $\mu_2$. Below the sample ($-t > z$), there is a non-superconducting substrate with a permeability $\mu_3$. The radius of the field coil and the pickup loop are $r$ and $r'$, respectively. By solving Maxwell's equations and the London equation for the three regions in the limit of $r' \ll r$ ($r=0.79~\mu$m and $r'\sim0.1~\mu$m), one obtains the SQUID height dependence of the susceptibility $\chi(z)$ as 
\begin{eqnarray}
    \chi(z)/\phi_s &=& \int_0^\infty dx \, e^{-2x\bar{z}} xJ_1(x)\nonumber\\
    &&\left[\frac{-(\bar{q}+\bar{\mu}_2x)(\bar{\mu}_3\bar{q}-\bar{\mu}_2x)+e^{2\bar{q}\bar{t}}(\bar{q}-\bar{\mu}_2x)(\bar{\mu}_3\bar{q}+\bar{\mu}_2x)}{-(\bar{q}-\bar{\mu}_2x)(\bar{\mu}_3\bar{q}-\bar{\mu}_2x)+e^{2\bar{q}\bar{t}}(\bar{q}+\bar{\mu}_2x)(\bar{\mu}_3\bar{q}+\bar{\mu}_2x)}\right],
\end{eqnarray}
where $\phi_s = A\mu_0/2\Phi_0r$ is the self inductance between the field coil and pickup loop, $A$ is the effective area of the pickup loop, $\bar{z} = z/r$, $J_1$ is the Bessel function of first order, $\bar{t} = t/r$, $\bar{q} = \sqrt{x^2 + \bar{\mu}_2 (r/\lambda)^2}$, and $\bar{\mu}_2=\mu_2/\mu_0$. For the bulk sample on a copper substrate ($\bar{t} \gg 1, \bar{\mu}_3 = 1$), the observed height dependence only depends on $\lambda$, $\mu_2$ and the SQUID structure. The observed susceptibility, which is normalized by the mutual inductance 55~$\Phi_0$/A at far from the sample surface, at different heights allowed us to estimate the local $\lambda$ by fitting to Eq.~(S4) with $\bar{t}=10$, $\bar{\mu}_2=1.03$, and $\bar{\mu}_3 = 1$. The permeability $\bar{\mu}_2$ is estimated by fitting the height dependence of the paramagnetic susceptibility at $T>T_c$. The scan height $z$ is determined as $z=z_{cal}(V-V_0)$, where $z_{cal}=0.5~\mu$m/V is estimated by fitting the height dependence of the paramagnetic susceptibility, $V_0=400$~nm is the distance between the pickup loop's center and the sample surface when the SQUID tip touches the surface ($V=0$), determined by optical measurements at room temperature,  $V=0$ is determined by detecting a kink in capacitance measurements. The error value of $z$ causes the biggest error for determining $\lambda(0)$. The errors in determining $V=0$ and $V_c$ are $\sim\pm$50~nm and $\sim\pm$100~nm, respectively. Error-bars in Figs. 3(c) and 4(a) are calculated by using $\pm$100~nm uncertainty in $z$.
Note that we used $V_0=500$~nm for $x=$0.018, 0.023, and 0.029 because the SQUID tip inadvertently picked up a small particle while measuring the $x=0.018$ sample, prior to measuring the other two. An estimate of 100~nm was determined by comparing the susceptibility amplitude of samples with particles to those without (data not shown).

\section{Fitting the temperature-dependent superfluid density}
We consider the 2D ellipsoid form of the superconducting gap as the anisotropic $s$-wave pairing symmetry. 2D ellipsoid form is expressed by $(x^2/a^2) + (y^2/b^2) = 1$ where $x = a\cos{\phi}$ and $y = b\sin{\phi}$. In the polar coordinate, this form is expressed by $r^2 = x^2 + y^2 = a^2\left[ 1 - \left( 1 - (b/a)^2 \right)\sin^2{\phi} \right]$. Thus we introduce the anisotropic $s$-wave symmetry model as $\Delta(T,{\bold k}) = \Delta_a(T)\times \sqrt{1-[1-(\Delta_c(0)/\Delta_a(0))^2]\sin^2{\phi}}=\Delta_0(T) \times g(\phi,\varepsilon)$, where $\varepsilon = 1-(\Delta_c(0)/\Delta_a(0))^2$. 

To fit our experimental results of the temperature-dependent superfluid density, for simple estimate, we use the approximate formula of 
\begin{equation}
\Delta'_0(T) = \Delta_0(T)/k_{\rm B}T_{\rm c} = \Delta'_0(0)\tanh(\pi\sqrt{a(T_{\rm c}-T)/T}/\Delta'_0(0))\label{eq:gap}
\end{equation}
with fitting parameters $a$ and $\Delta'_0$~\cite{Gross1986,Prozorov2006}.


The temperature-dependent superfluid density is expressed by the semi-classical model as Eq.~(2) in the main paper. To fit the experimentally obtained superfluid density $n^{exp}_s(T)$ by using the simulation model $n^{sim}_s(T,a,\varepsilon,\Delta_a(0),\lambda(0))$, we calculate the error square sum $\Xi=\sum_i[n_s(T_i) - n^{sim}_s(T_i,a,\varepsilon,\Delta_a(0),\lambda(0))]^2$. Figures S4 estimate for the uncertainties in the parameters $a, \Delta_c(0)/\Delta_a(0), \Delta_a(0)/k_{\rm B}T_{\rm c},$ and $\lambda(0)$~\cite{Kirtley2012}. The errors in parameters are determined by the double amount of the global minimum $\Xi_{\rm min}$. The errors in $\lambda(0)$ are estimated as less than $\pm40$ nm, which is much shorter than the systematic errors due to uncertainty in determining the scan height, $\pm100$~nm, that is used in Fig.~3(c).

\begin{figure}[htb]
\includegraphics*[width=14cm]{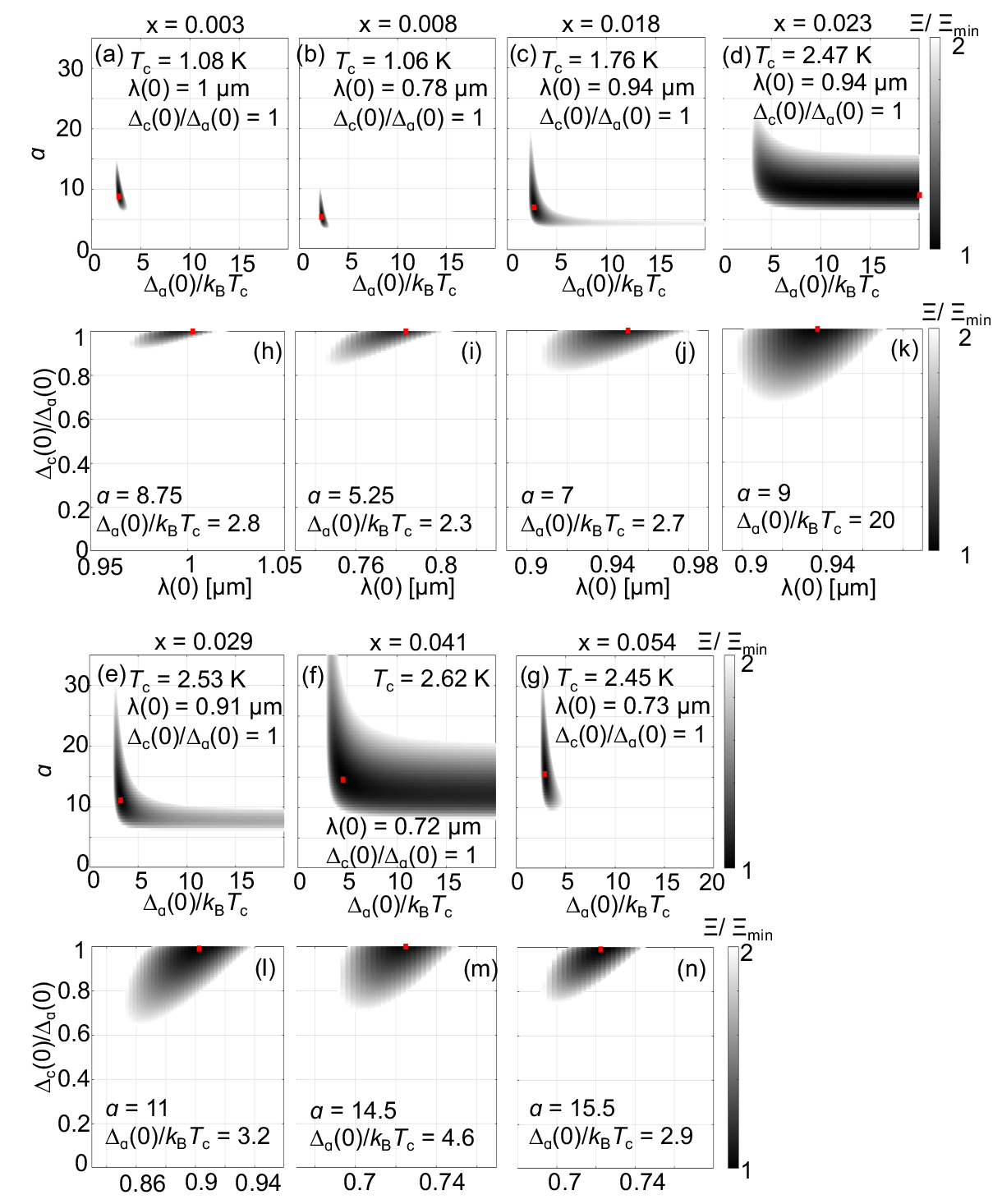}
\caption{\label{fig:ani-gap} The error square sum $\Xi$ normalized by the global minimum $\Xi_{\rm min}$ are plotted for (a-g) fitting parameter $a$ and the coupling amplitude $\Delta_a(0)/k_BT_c$ 
 or (h-n) anisotropy $\Delta_c(0)/\Delta_a(0)$ and $\lambda(0)$ in Eq.~(\ref{eq:gap}) for each experimental results. Red dots are the global minimum points. Error bars in Figs. 4(b,c) are determined by the range of $\Xi_{\rm min}<2$.}
\end{figure}

\begin{figure}[htb]
\includegraphics*[width=14cm]{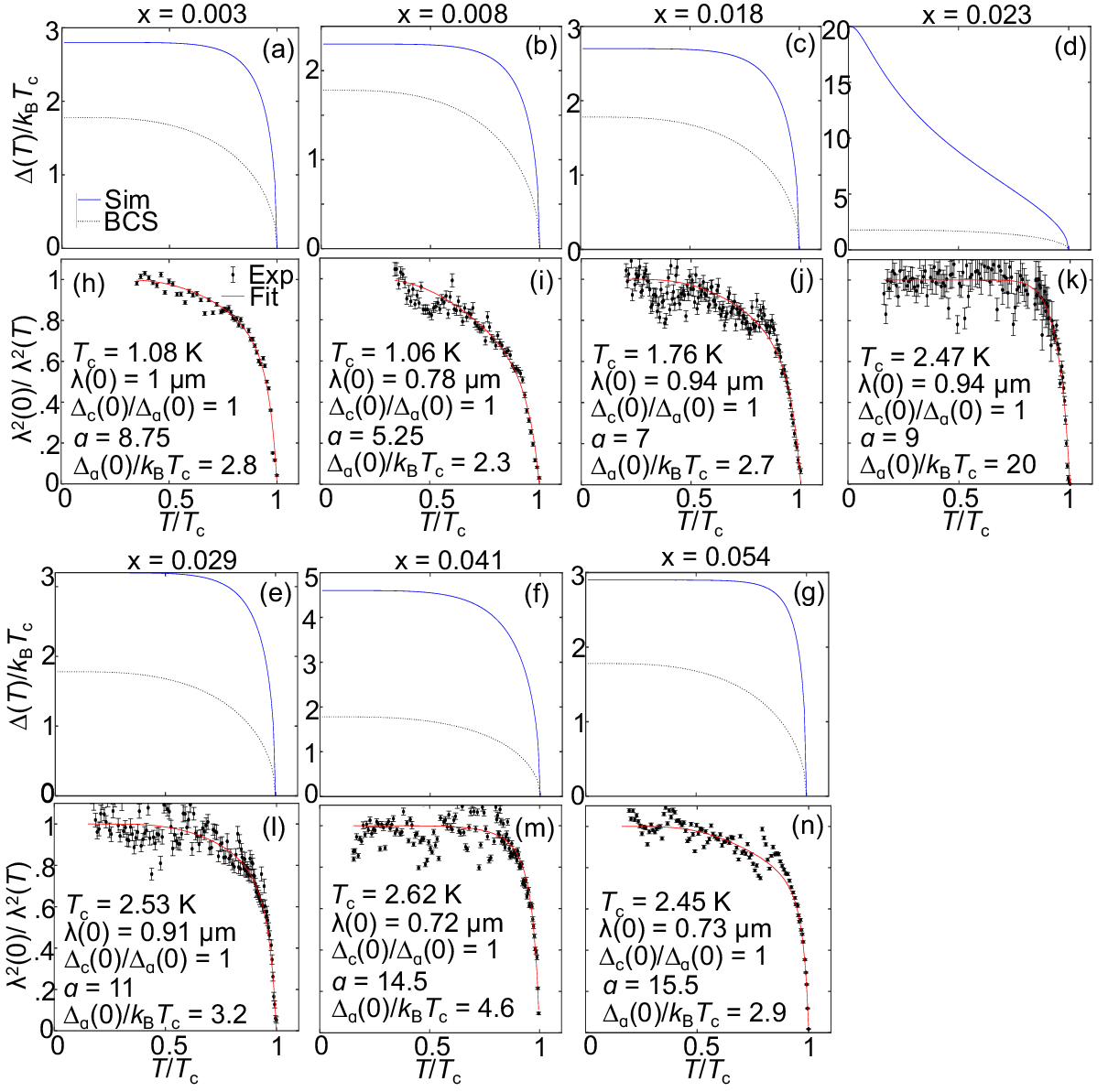}
\caption{\label{fig:tc-lam0} The best fitting results for all measured superfluid density. (a-g) Temperature-dependent superconducting gap with best fitting parameters of $a, \Delta_a(0)/k_BT_c, \Delta_c(0)/\Delta_a(0)$.}
\end{figure}

\clearpage
\section{Computing the superfluid density from quantum phase
fluctuations}
\subsection{The problem}

We imagine that it is only quantum and thermal phase fluctuations that determine the evolution of the superfluid density, i.e.~local pairing is assumed to exist over a broader range of $T$. The simplest such model is the quantum rotor model, which we will adopt for simplicity---although more realistic models can be treated in similar fashion.

We thus consider the Hamiltonian
\be
\label{eq:ham}
H= \sum_j \frac {n_j^2} {2C} - J\sum_{\langle i,j\rangle} \cos(\theta_i - \theta_j) ,
\ee
where $n_j$ is the number of Cooper pairs on site $j$ and satisfies the commutation relations:
\be
\left[ n_i,n_j\right] =\left [e^{i\theta_i},e^{i\theta_j}\right]=0
\ee
and
\be
\left[n_i,e^{i\theta_j}\right] = \delta_{ij} \, e^{i\theta_j} .
\ee

This can be expressed in terms of the imaginary time effective action

\be
S=\int_0^\beta d\tau \left\{ \frac C 2 \sum_j \dot\theta_j^2 - J\sum_{\langle i,j\rangle} \cos(\theta_i - \theta_j) \right\} .
\ee

\subsection{Variational solution}
We treat the problem using the same variational method as used (for a related problem) in Ref.~\onlinecite{sudip}. We introduce a variational effective action,
\be
S_{\text{tr}}=\int_0^\beta d\tau \left\{ \frac C 2 \sum_j \dot\theta_j^2 + \frac{\mu} 2 \sum_{\langle i,j\rangle} (\theta_i - \theta_j)^2 \right\} ,
\ee
and choose $\mu$ to minimize the variational free energy.  
The trial action is quadratic and can be diagonalized by inserting Fourier expansions of the fields (here $\omega_n = 2\pi n/\beta$ is a bosonic Matsubara frequency):
\be
\theta_j(\tau) = \frac{1}{\beta} \sum_n \int \frac{d^dk}{(2\pi)^d} \ e^{i\omega_n \tau - i \vec k \cdot \vec r_j} \, \theta_{\vec k,n} .
\ee
Since $\theta_j(\tau)$ is real, $\theta_{\vec k, n} = (\theta_{-\vec k, -n})^*$.

(Note that here we have completely ignored the fact that $\theta$ is an angular variable, i.e.~we have neglected vortices.
We should in principle account for this fact by including linear terms of the form $2\pi m \tau/\beta$ in the Fourier expansion of $\theta(\tau)$, where $m$ is an integer winding number. However, this would complicate the analysis, and presumably only alter the results significantly when $\mu$ is very small---perhaps small enough to be pre-empted by a first-order transition to the disordered state. This justifies the present ``spin-wave'' approximation in which we ignore the periodicity of $\theta$.)

Inserting the Fourier expansion of $\theta_j(\tau)$, the variational effective action becomes
\be
S_{\text{tr}} = \frac{C}{2\beta} \sum_n \int \! \frac{d^dk}{(2\pi)^d} \left( \omega_n^2 + \Omega_{\vec k}^2 \right) \abs{\theta_{\vec k,n}}^2 ,
\ee
with
\be
\Omega_{\vec k}^2 = \frac{\mu}{2C} \sum_{j = \text{n.n.} \, i} \abs{1 - e^{i \vec k \cdot \vec R_{ij}}}^2
= \frac{\mu}{C} \sum_{a=1}^d \sin^2(k_a/2) ,
\ee
where the second equality is for the $d$-dimensional hypercubic lattice.

Since the trial action is quadratic and diagonal in $\theta_{\vec k,n}$, one has
\be
\langle \theta_{\vec k,n} \theta^*_{\vec k',m} \rangle_{\text{tr}} = (2\pi)^d \delta(\vec k - \vec k') \, \delta_{n,m} \ \frac{\beta}{C(\omega_n^2 + \Omega_{\vec k}^2)} .
\ee
It follows that
\be
\langle (\theta_i-\theta_j)^2 \rangle_{\text{tr}} 
= \frac{1}{\beta C} \sum_n \int \! \frac{d^dk}{(2\pi)^d} \ \frac{1}{\omega_n^2 + \Omega_{\vec k}^2} \abs{1 - e^{i \vec k \cdot \vec R_{ij}}}^2 .
\ee
Performing the Matsubara sum over $n$ yields
\be
\langle (\theta_i-\theta_j)^2 \rangle_{\text{tr}} 
= \int \! \frac{d^dk}{(2\pi)^d} \left[ \frac{2f(\Omega_{\vec k}) + 1}{2 C \Omega_{\vec k}} \right] \abs{1 - e^{i \vec k \cdot \vec R_{ij}}}^2 ,
\ee
where $f(\omega) = (e^{\beta\omega} -1)^{-1}$  is the Bose occupation factor.
Moreover, because the trial action is quadratic,
\be
\langle\exp[i(\theta_i-\theta_j)]\rangle_{\text{tr}} = \exp\!\left[-\frac 1 2 \langle[\theta_i-\theta_j]^2\rangle_{\text{tr}} \right] .
\ee

We now use the variational inequality
\be
F \leq F_{\text{tr}} + \frac{1}{\beta} \langle S - S_{\text{tr}} \rangle_{\text{tr}} .
\ee
The right-hand side is (using the formulae above)
\be
F_{\text{tr}} - \sum_{\langle i,j\rangle} \left\{ J e^{-\frac{1}{2} \langle[\theta_i-\theta_j]^2\rangle_{\text{tr}}} + \frac{\mu}{2}  \langle (\theta_i - \theta_j)^2 \rangle_{\text{tr}} \right\} .
\ee
The derivative of this expression with respect to $\mu$ should vanish.
Using the fact that $\partial F_{\text{tr}}/ \partial\mu = \frac{1}{2} \sum_{\langle i,j\rangle} \langle (\theta_i - \theta_j)^2 \rangle_{\text{tr}}$, we obtain the self-consistency condition for $\mu$:
\be
\mu = J  \exp\left[-\frac{1}{2} \langle[\theta_i-\theta_j]^2\rangle_{\text{tr}} \right] .
\ee

The trial action $S_{\text{tr}}$ describes a collection of harmonic oscillators with mass $C$ and frequencies $\Omega_{\vec k}$ at temperature $1/\beta$. Thus, the trial free energy per site is
\be
\frac{F_{\text{tr}}}{V} = \int \! \frac{d^dk}{(2\pi)^d} \left\{ \frac{1}{2} \Omega_{\vec k} + \frac{1}{\beta} \log\!\left(1 - e^{-\beta \Omega_{\vec k}}\right) \right\} .
\ee
It follows that the variational inequality is $F \leq \tilde{F}_{\text{tr}}$, where
\be
\frac{\tilde{F}_{\text{tr}}}{V} = \int \! \frac{d^dk}{(2\pi)^d} \left\{ \frac{1}{2} \Omega_{\vec k} + \frac{1}{\beta} \log\!\left(1 - e^{-\beta \Omega_{\vec k}}\right) \right\}
- \frac{1}{2} \sum_{j = \text{n.n.} \, i} \left\{ J e^{-\frac{1}{2} \langle[\theta_i-\theta_j]^2\rangle_{\text{tr}}} + \frac{\mu}{2}  \langle (\theta_i - \theta_j)^2 \rangle_{\text{tr}} \right\} .
\ee
Inserting the self-consistency condition for $\mu$, this reduces to
\be
\frac{\tilde{F}_{\text{tr}}}{V} = \int \! \frac{d^dk}{(2\pi)^d} \left\{ \frac{1}{2} \Omega_{\vec k} + \frac{1}{\beta} \log\!\left(1 - e^{-\beta \Omega_{\vec k}}\right) \right\}
- \frac{\mu}{2} \sum_{j = \text{n.n.} \, i} \left[ 1 + \frac{1}{2} \langle (\theta_i - \theta_j)^2 \rangle_{\text{tr}} \right] .
\ee
The right hand side should be compared with the same quantity computed in the disordered state.  For the disordered state we take the free action, i.e. $S_{\text{free}}$ is the same as $S_{\text{tr}}$ but with $\mu=0$.
The corresponding free energy per site is
\be
\frac{F_{\text{free}}}{V} = - \frac{1}{\beta} \log[ \Theta(e^{-\beta/2C}) ] ,
\ee
where 
\be
\Theta(x) \equiv  \sum_{n=-\infty}^\infty x^{n^2} .
\ee
In addition, since the sites are decoupled in the free limit, $\langle S - S_{\text{free}} \rangle_{\text{free}} = 0$.

\subsection{Summary}

The self-consistency condition for $\mu$ can be written as
\be
\label{eq:cons}
\mu = J \exp\left[- \ \frac{\beta}{C} \ W\big(\beta \sqrt{\mu/C}\big) \right] ,
\ee
where
\be
W(z) = \frac{1}{4d} \int \! \frac{d^dk}{(2\pi)^d} \ \frac{1}{z} \, [g(\vec k)]^{1/2} \coth\!\left( \frac{z}{2} \, [g(\vec k)]^{1/2} \right)
\ee
and $g(\vec k)$ is the structure factor
\be
g(\vec k) = \sum_{a=1}^d \sin^2(k_a/2) .
\ee
Eq.~(\ref{eq:cons}) must be solved numerically for $\mu$.

The variational free energy per site can be written as
\be
\frac{\tilde{F}_{\text{tr}}}{V} = \frac{1}{\beta} \ Y\big(\beta \sqrt{\mu/C}\big)
- \mu d ,
\ee
where
\be
Y(z) = \int \! \frac{d^dk}{(2\pi)^d} \left\{ \frac{z}{2} \, [g(\vec k)]^{1/2} \left[ 1 - \frac{1}{2} \coth\!\left( \frac{z}{2} \, [g(\vec k)]^{1/2} \right) \right]
+ \log\!\left(1 - e^{-z [g(\vec k)]^{1/2}}\right) \right\} .
\ee
This should be compared to the free energy per site of the disordered state:
\be
\frac{F_{\text{free}}}{V} = - \frac{1}{\beta} \log[ \Theta(e^{-\beta/2C})], \qquad
\Theta(x) \equiv  \sum_{n=-\infty}^\infty x^{n^2} .
\ee

\begin{figure}[h]
\includegraphics*[width=12cm]{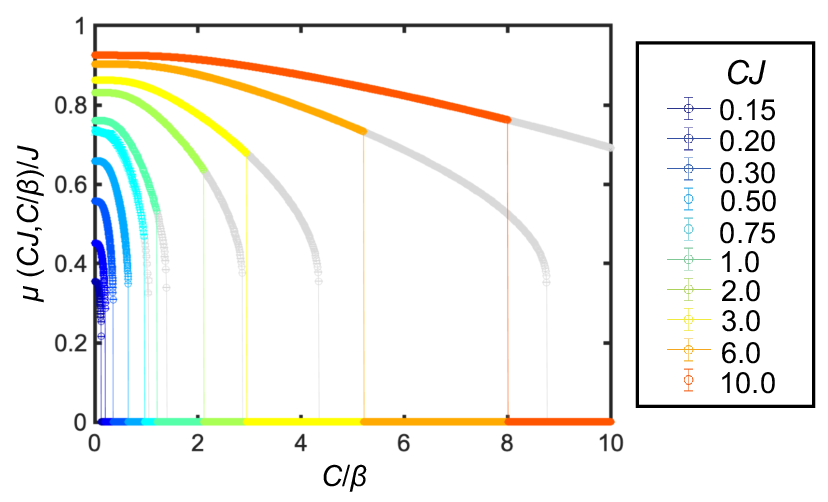}
\caption{\label{fig:Tdepmu} Temperature dependence of the calculated normalized superfluid density $\mu(CJ,C/\beta)/J$ in the toy model of Eq.~(\ref{eq:ham}). The grey data are the calculated results where $F_{\text{free}}<\tilde{F}_{\text{tr}}$.}
\end{figure}

\begin{figure}[hb]
\includegraphics*[width=17cm]{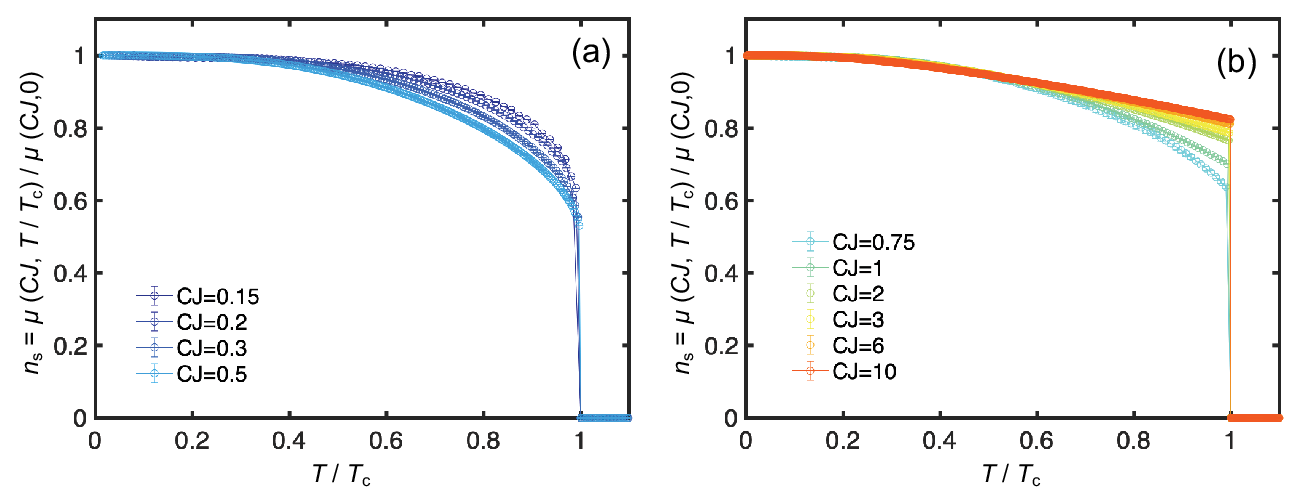}
\caption{\label{fig:TdepSF} Normalized temperature dependence of the superfluid density $\mu(CJ,T/T_c)/\mu(CJ,0)$ in the toy model of Eq.~(\ref{eq:ham}). (a) With weak coupling constants $CJ<0.55$, the ordered state is stable ($F_{free}>\tilde{F}_{tr}$). (b) With strong coupling constants $CJ>0.55$, the superfluid density has a linear temperature dependence near $T_c$.}
\end{figure}

The calculation results are plotted as a function of $C/\beta$ with coupling constants $CJ$ in Fig \ref{fig:Tdepmu}, where the transition temperature $C\beta_c \propto T_c$ was determined by the self-consistent solution of Eq.~(\ref{eq:cons}) with weak coupling constants $CJ<0.55$ or suppressed due to $F_{\text{free}}<\tilde{F}_{\text{tr}}$ with strong coupling constants $CJ>0.55$. From these results, the normalized temperature ($T/T_c=\beta_c/\beta$) dependence of the normalized superfluid density ($n_s = \mu(CJ,T/T_c)/\mu(CJ,0)$) are obtained as shown in Fig. \ref{fig:TdepSF}.

\end{document}